\documentclass{iopart}

\usepackage{amssymb}
\usepackage{graphicx}

\begin{document}


 \title{Charm Production in {\sc dpmjet}}
 \author{P Berghaus$^1$, T Montaruli$^{1,2}$ and J Ranft$^3$}
 \address{$^1$ University of Wisconsin - Madison, WI 53706, USA}
 \address{$^2$ on leave from Universit\`a di Bari, 70126, Italy}
 \address{$^3$ University of Siegen, D-57068, Germany}
 \eads{\mailto{berghaus@icecube.wisc.edu}, \mailto{montaruli@ba.infn.it} and \mailto{ran@cern.ch}}

\begin{abstract}

In this work, charm production in the {\sc dpmjet} hadronic jet simulation is compared to experimental data. Since the major application of {\sc dpmjet} is the simulation of cosmic ray-induced air showers, the version of the code integrated in the CORSIKA simulation package has been used for the comparison. Wherever necessary, adjustments have been made to improve agreement between simulation and data. With the availability of new muon/neutrino detectors that combine a large fiducial volume with large amounts of shielding, investigation of prompt muons and neutrinos from cosmic ray interactions will be feasible for the first time. Furthermore, above $\gtrsim 100$ TeV charmed particle decay becomes the dominant background for diffuse extraterrestrial neutrino flux searches. A reliable method to simulate charm production in high-energy proton-nucleon interactions is therefore required.

\end{abstract}





\section{Introduction}

During recent years neutrino detectors, notably IceCube and ANTARES, as well as large air showers arrays, such as the Pierre Auger Observatory and the Telescope Array, have been advancing or completing their construction. These large detectors are sensitive to an energy region in which contributions from prompt charm decays in cosmic ray showers cannot be neglected and may constitute an interesting signal as well as a significant background. In the case of searches for diffuse fluxes of astrophysical neutrinos in neutrino telescopes, the signal must to be separated at high energies from the background of atmospheric neutrinos. 

Atmospheric neutrinos from $\pi$ and K meson decays, whose spectrum approximately follows the power law $E^{-3.7}$ above 100GeV, can be rejected using cuts on energy-correlated variables. However, the spectrum of prompt neutrinos is expected to be harder and correspond to the primary cosmic ray spectrum, since charmed mesons tend to decay promptly without interacting in the atmosphere (see e.g. \cite{Gelmini:2002sw,Martin:2003us}). Consequently, this contribution needs to be known and separated from the signal, which is generally assumed to follow a power law of $\approx E^{-2}$, characteristic of Fermi acceleration processes \cite{diffuse}. It is therefore essential to achieve a precise understanding of the expected charm-induced lepton flux and angular distribution, in order to identify it in the data above 100-1PeV, where the contribution of the softer component from light meson decays becomes negligible. For this purpose, the entire air shower needs to be simulated. 

One of the most widely used cosmic ray shower simulation software packages in high-energy astrophysics is CORSIKA \cite{Heck:1998vt}. It offers the user the choice between several high-energy hadronic jet simulations but does not yet officially support propagation and decay of charmed particles. In order to remedy this situation, first the viability of one of the available models for simulation of charm has to be demonstrated.

The consistency of charm simulation with experiment was first verified in 1995 using the standalone version of {\sc dpmjet}\cite{Battistoni95b}. The best available measurements at the time were those from the LEBC hydrogen bubble chamber that took data at CERN and Fermilab during the late 1980s using proton beams of 400GeV and 800GeV, respectively \cite{AguilarBenitez:1988sb,Ammar:1988ta}. Since then, a significant amount of new experiments have taken place, most notably in fixed-target runs at Fermilab, but also using various types of colliders and most recently the HERA-B detector at DESY \cite{Abt07}. In this work, the version of {\sc dpmjet} integrated in CORSIKA was used, running in first-interaction mode. This was done to directly evaluate its applicability to air shower simulation and to ensure compatibility with the standalone version.

\section{Charm production in the Dual Parton Model}
\label{sec:5}

\subsection{The Dual Parton Model}

There are two mechanisms in any model for hadron production:
soft particle production and hard collisions. 
Both components are part of the {\sc dpmjet} hadron production
models used in this work.
 Soft hadron production cannot be rigorously derived from
QCD, the gauge field theory of strong interactions. Most theoretical
efforts use Regge theory for a systematic description of soft
hadron production. 
The so--called {\it hard}
component of particle production  can be derived from
perturbative QCD. QCD perturbation theory within the QCD
improved parton model can only be applied at moderately large transverse
momenta ($p_{\perp}$). One feature of soft hadron production at small $p_{\perp}$ values is represented by
exponentially decreasing transverse momentum distributions. The
QCD improved parton model predicts another component of the transverse
momentum distributions, decreasing like a power law, and thus less steeply than an exponential function, at large $p_{\perp}$.

The Dual Parton Model {\bf DPM} \cite{DPMREV92} and the closely related
Quark Gluon Strings Model {\bf QGSM} \cite{Kaidalov82a,Kaidalov82b}
construct hadron production from fragmenting strings.
Here we will  describe the
structure of this mechanism in hadron-hadron collisions.
Since it is not possible to describe soft component of 
hadron physics using perturbative
QCD, 't Hooft introduced a new expansion parameter \cite{Hooft74}.
First, QCD is generalized from the gauge group SU(3) to SU($N_c$) with
$N_c$ representing the number of quark ``colours''. The idea was to use
$1/N_c$ as an artificial expansion parameter, and later set
$N_c$ = 3 for physical applications. One then finds
that the Feynman graphs can be characterized by a two-dimensional surface, and
the Feynman expansion can be regrouped as a sum over surface
topologies. The leading order corresponds to planar graphs, the next order for 2-to-2 amplitudes involves a cylindrical topology.  
It has been shown \cite{Ciafaloni75} that there is a one-to-one
correspondence between the terms of this topological expansion and the
terms of Reggeon Field Theory {\bf RFT}\cite{Abramovski73}.
The second theoretical concept entering DPM is duality
\cite{Jacob74} and the 
Dual Topological Unitarization {\bf DTU}\cite{Chan75}.
The third ingredient of DPM is the coloured parton model.

 The leading contribution to soft hadron production in the DPM   
corresponds to the production of two chains. The two chains
arise from the unitarity cut of the Pomeron exchange diagram. Located at the end
of either chain are the valence quarks and diquarks of the two protons. The valence diquarks have anti-triplet colour, while the
colliding protons as well as the two produced chains are colour neutral.
Particle production occurs via fragmentation of the two
quark-diquark chains. In order to arrive at a quantitative model, we have
to specify the probability $\rho_1 (x)$ that the interaction separates
the proton into a valence quark with momentum fraction $x$ and a
valence diquark with the remaining momentum fraction $(1 - x)$. These
longitudinal momentum fractions are given by Regge asymptotics. Valence quarks in baryons as well as in mesons follow
$x^{-1/2}$, valence diquarks in baryons $(1 -
x)^{3/2}$ and sea quarks $x^{-1}$ distributions.
Combining these, we obtain for the distribution in a proton 
\begin{equation}
 \rho^p_1 = c^p_1x^{-1/2}(1 - x)^{3/2},
 \label{rhobaryon}
\end{equation}
and in a meson (valence quarks follow the same distribution as
their corresponding antiparticles) 
\begin{equation}
 \rho^m_1 = c^m_1x^{-1/2}(1 - x)^{-1/2}.
 \label{rhomeson}
\end{equation}
where the
coefficients $c^p_1$ and $c^m_1$ normalize the distributions to unity.

This two-chain model already describes hadron production at low energies quite successfully. Many Monte Carlo models therefore
started with two-chain models.
However, at higher energies more and more discrepancies between the
model and experimental data appear. One example is that the measured
transverse momentum distributions can only be explained by including a
hard component in the hadron production.
Also, from a theoretical point of view, the pomeron total cross section
(using the so-called supercritical pomeron) violates the unitarity principle.
Therefore, the model has to be unitarized.

The {\sc dtujet}\cite{Aurenche92a} and {\sc
phojet}\cite{Engel95a,Engel95d}
Monte Carlo models use an eikonal unitarization scheme. 
 {\sc dtujet} and {\sc phojet}, respectively, 
 are used for the elementary
 interactions in {\sc dpmjet}-II and {\sc dpmjet}-III. Both
 models have multiple soft and hard chains as demanded by the
 unitarization method.

{\sc dpmjet}-II \cite{DPMJETII} 
and {\sc dpmjet}-III \cite{Roesler20002}
use the Glauber model to
describe hadron--nucleus and nucleus--nucleus collisions. Here, we
will not present any details about the Glauber model and
refer the reader instead to the original publications cited above.

\subsection{Charm production in {\sc dpmjet} and CORSIKA}
\label{bugsec}
The paper by Battistoni et al. \cite{Battistoni95b} gives the first description of DPM charm production.  They used a version interfaced to the cosmic ray
cascade code HEMAS \cite{HEMAS}, which has since then been phased out. There, DPM was implemented in form
of the Monte Carlo event generator {\sc dpmjet}-II \cite{Roesler20002}, which itself is based on earlier work \cite{Shabelski93,Kaidalov86,Derujula92}. 

In \cite{Battistoni95b}, hard charm production was tested against NLO perturbative calculations
\cite{Greco94} and differential charm production cross sections
were compared to experimental data, mainly from the LEBC-EHS and
LEBC-MPS collaborations \cite{AguilarBenitez:1988sb,Ammar:1988ta}.
Hadronic jet simulation with {\sc dpmjet}-II is available in the CORSIKA cosmic ray transport
code \cite{Heck:1998vt}. However, the official CORSIKA version at the time {\sc dpmjet}-II was
implemented had no provision for the propagation of charmed hadrons. Instead, all charmed hadrons not decaying within the hadronic jet simulation itself were handled by substituting strange quarks for charm. Charm production from {\sc dpmjet}-II and other
jet simulations is expected to become available in the official CORSIKA release \cite{Heck_private} in the near future. Earlier, an unofficial modified version was provided to the authors \cite{Meli}, in which charm was
no longer supressed. Our work is mostly based on this version, in which however one crucial parameter was erroneously changed with respect to the original {\sc dpmjet}. The comparisons with experiment shown below demonstrate both the influence of correcting this parameter and the modifications made subsequently to the main {\sc dpmjet} code (see Figs. \ref{selex_fig} and \ref{hera_fig}).

{\sc dpmjet}-II is also integrated in the FLUKA hadron cascade code
\cite{FLUKA}.  We are not aware of any charm calculations 
using FLUKA-{\sc dpmjet}-II, but this might change soon. All modifications to
charm production described in this paper are similarly
applicable to FLUKA.

\subsection{Charm production in elementary hadronic collisions}


In {\sc dpmjet}-II there are three different mechanisms by which production of heavy flavors like charm can occur: (i) charmed quarks produced
at the ends of hard and semihard chains (minijets), (ii) charmed
quarks produced at the ends of soft sea chains and (iii) charm
production inside the chain decay. Only the first mechanism (i)
is founded on the solid theoretical basis of perturbative
QCD, the other two mechanisms are phenomenological models
used for charm and prompt muon production within cosmic ray-induced
cascades

\subsection{Charm production at the ends of hard and semihard
chains (minijets)}

Regarding this component, nothing in {\sc dpmjet}-II was changed with respect
to the older version described in \cite{Battistoni95b}. Therefore, we do not need
to repeat the details given there. 
However, it should be mentioned that in that work
a detailed comparison was presented between the results of the {\sc
dpmjet}-II Monte Carlo
at large transverse momenta and the corresponding NLO QCD
calculations, showing excellent agreement between the two. For hadronic and
nuclear collisions at laboratory energies beyond the TeV range, this is the dominant mechanism of charm
production.

 \subsection{Charm production at the ends of soft sea chains}

In {\sc dpmjet}, charm production at the end of soft sea chains
is considered as the non-perturbative limit of minijets. The
parton transverse momentum distributions of minijets and
soft sea chains are joined smoothly at the threshold transverse
momentum between the two. A certain fraction of the soft sea
chain ends carry heavy flavors. At high energies, the
probability $P_{c \bar c}$ for production of a $c \bar c$ sea quark pair
approaches the corresponding value for semi-hard chains, while at low energies
$P_{c \bar c}$ decreases as required by phenomenological considerations.
 {\sc dpmjet} uses
\begin{equation}
P_{c \bar c} = C\int^{E_R}_{m_q} 2E_{\perp}e^{-b_c E_{\perp}}
dE_{\perp}
\end{equation}
with
\begin{equation}
b_c = b + 1.3 - \log10 (\frac{E_{CM}}{1GeV}).
\end{equation}
An additional constraint is given by the requirement that $P_{c \bar c}$ should not
exceed the corresponding probability for minijets.

 \subsection{Charm production inside the soft chain
 fragmentation}
\label{sec-selex}
{\sc dpmjet}--II uses the Lund model
{\sc pythia} for chain fragmentation \cite{Sjostrand01a}. Within this model, a
quark-antiquark pair $q_i \bar q_i$ leading to string
breakup is produced via quantum mechanical tunneling. The
tunneling probability can be worked out as function of the
transverse mass $m_{\perp}$ of the pair
\cite{Sjostrand01a}, with the result that $c \bar c$ pair production is highly
supressed: $u \bar u$:$d \bar d$:$s \bar s$:$c \bar c$ = 1 : 1 :
0.3 : 10${}^{-11}$. There are other models for string
fragmentation with larger $c \bar c$ production probabilities, but none in which charm production inside the soft chain could not be neglected.

In {\sc dpmjet}--II the picture remains the same, with one exception: the $c
\bar c$ probability near a valence diquark at the end of a
diquark-quark or diquark-antidiquark chain is not negligible. This effect was demonstated in hadroproduction by fixed target experiments such as SELEX \cite{SELEX,SELEX01}. The leading particle shares a valence quark with the incident hadron, while the non-leading one does not. Therefore leading particles are copiously produced at large $x_F$ in the forward region of the incident hadron, resulting in an asymmetry between leading and non-leading particles. This phenomenon is referred to as {\it leading quark effect} \cite{Tashiro:2003pd}.
SELEX finds in the fragmentation region of
protons and $\Sigma^-$ hyperons $\Lambda_c^+$ charmed hyperons
with a comparatively flat Feynman-x ($x_F$) distribution. In a fit to $(1-|x_F|)^{\alpha}$, 
$\alpha$ is about 2.5.
Without a new mechanism, {\sc dpmjet} generates this distribution with an
$\alpha$ of about 7. Better agreement with SELEX can be achieved by
introducing, adjacent to a diquark, $c \bar c$ pairs with
increased probability. In order to retain the standard {\sc pythia} code, this change was done in a special routine of {\sc dpmjet}-II. The comparison of SELEX data and the modified version of {\sc dpmjet} is shown in section 3.3.

\section{Comparison with Experimental Data}

\subsection{Collision Scaling of Charm Production}

In {\sc dpmjet}, nuclear collisions are treated with a Monte-Carlo formulation of the Glauber model
 \cite{DPMJETII,Roesler20002}. This model contains two important numbers: the number of participants
 $N_{part}$ and the total number of collisions $N_{coll}$. Soft
 particle production scales rather well with $N_{part}$, whereas hard
 particle production (for instance at large $p_{\perp}$) scales
 with $N_{coll}$. The latter phenomenon is commonly referred to as {\it collision scaling}. All charm
 production, because of the large mass difference of the quarks
 involved, is expected to follow collision scaling.

 This behaviour can be reproduced in experimental
 data. Fermilab experiment E789 \cite{Leitch94} measured
 neutral D-meson  
 production in p-Be and p-Au collisions, finding the
 production cross section to behave like
\begin{equation}
\sigma(A) = \sigma_0 A^{\alpha}
\end{equation}
 with $\alpha = 1.02 \pm -.03 \pm 0.02 $  
 and the ratio
\begin{equation}
R = \frac{\sigma(Au)/197}{\sigma(Be)/9} = 1.06 \pm 0.11 \pm
0.07.
\end{equation}
The consistency of these values with unity corresponds to collision scaling.

\begin{table}
\caption{\label{tab:1}Comparison of collision scaling in {\sc dpmjet}--II
with the results from HERA-B \cite{Abt07}. (all cross sections in
mbarn)}
\begin{indented}
%
%
\item[]\begin{tabular}{@{}llll}
\br
  & p-p & p-Nitrogen & HERA--B  \\
\mr
 $D^+$ $\sigma_{pN}$ & 0.0112 & 0.0144 & 0.0202$\pm$0.0022$\pm$0.0024$\pm$0.0018 \\
 $D^0$ $\sigma_{pN}$ & 0.0350 & 0.0445 & 0.0487$\pm$0.0047$\pm$0.0049$\pm$0.0044 \\
 $D_s$ $\sigma_{pN}$ & 0.0051 & 0.0068 & 0.0185$\pm$0.0064$\pm$0.0037$\pm$0.0017 \\
\br
\end{tabular}
\end{indented}
\end{table}

The HERA-B Collaboration \cite{Abt07} measured $D^0$, $D^+$,
$D^+_s$ and $D^{*+}$ in p-C, p-Ti and p-W collisions.
They parametrize the production cross sections as
\begin{equation}
\sigma_{p-A} = \sigma_{p-N} A^{\alpha}
\end{equation}
and obtain $\alpha = 0.99 \pm 0.04 \pm 0.03$ in agreement with collision scaling.

{\sc dpmjet}--II in its original form included charm production in which collision scaling was represented rather poorly. This has been
corrected in the version used in this paper.

In Table \ref{tab:1} we compare the $ \sigma_{p-N}$ cross
sections obtained from {\sc dpmjet}--II in p-p and p-Nitrogen
collisions with the ones obtained by the HERA--B collaboration. 
Taking into account experimental errors (as given in the table), we
find good agreement, meaning that collision scaling as simulated in {\sc
dpmjet}-II corresponds to that found experimentally.

\begin{figure}
\begin{center}
\includegraphics[width=4in]{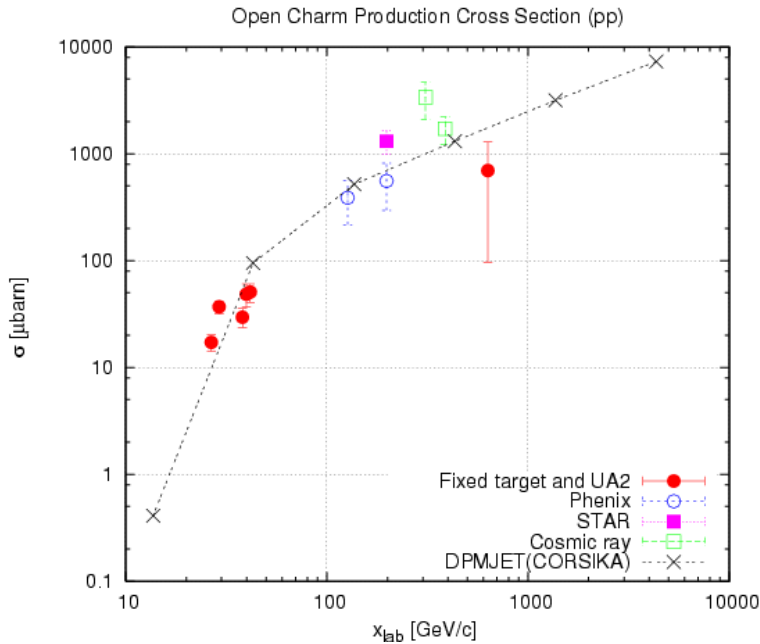}
\caption{Total open charm production cross section measurements, taken from \cite{Merino:2007rg} and references therein, represented by markers. The corresponding values from {\sc dpmjet} simulation are indicated by the dotted line.}
\label{total_csec_fig} 
\end{center}
\end{figure} 

\subsection{Total Charm Production Cross Section}

Figure \ref{total_csec_fig} shows the total open charm production cross section as measured, along with the corresponding values from {\sc dpmjet}. It can be seen that the simulation reproduces the actual values reasonably well, considering that the errors on the experimental values are quite large.


\subsection{Differential Cross Sections}

\begin{figure}
\includegraphics[width=3in]{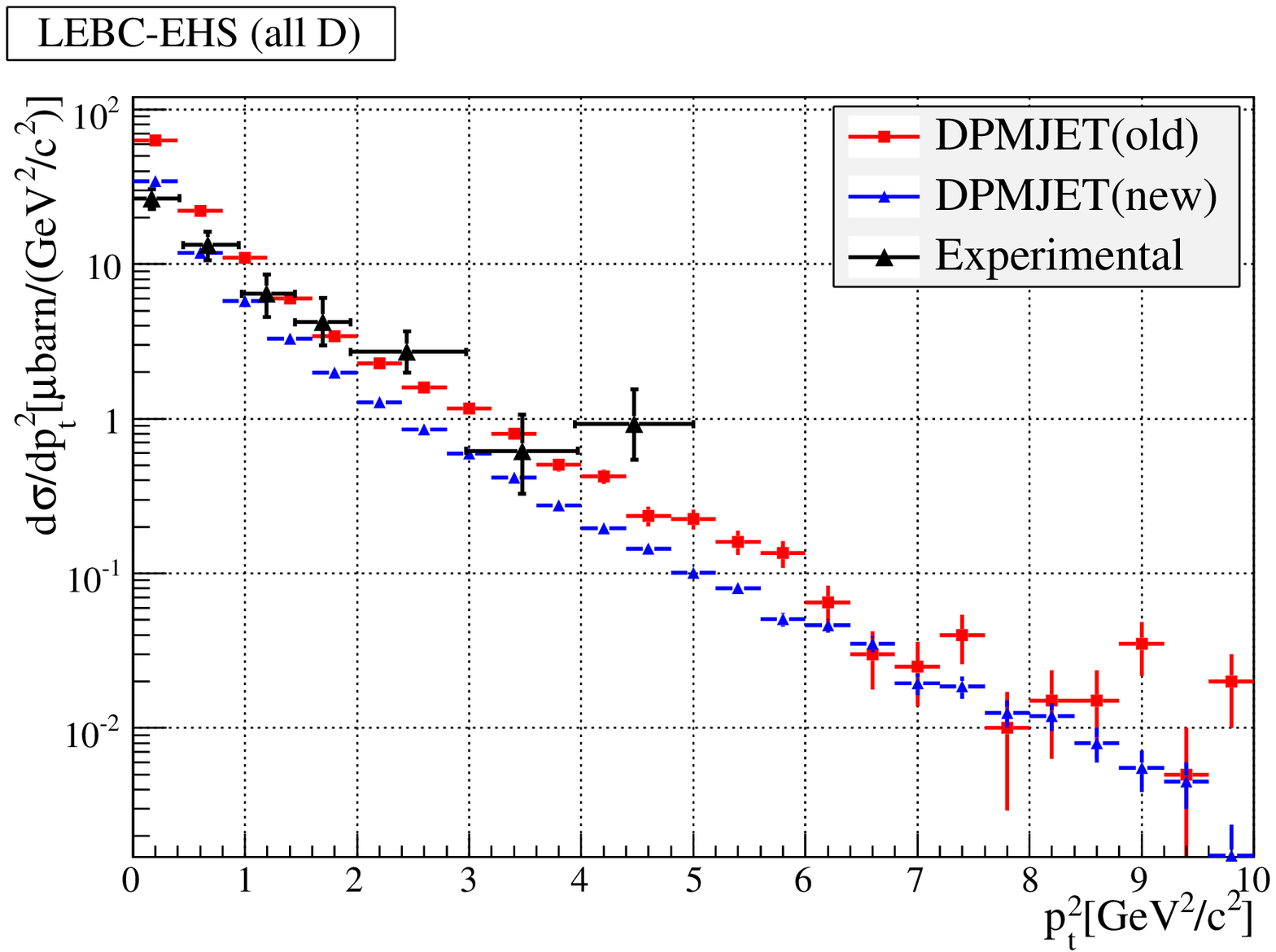}
\includegraphics[width=3in]{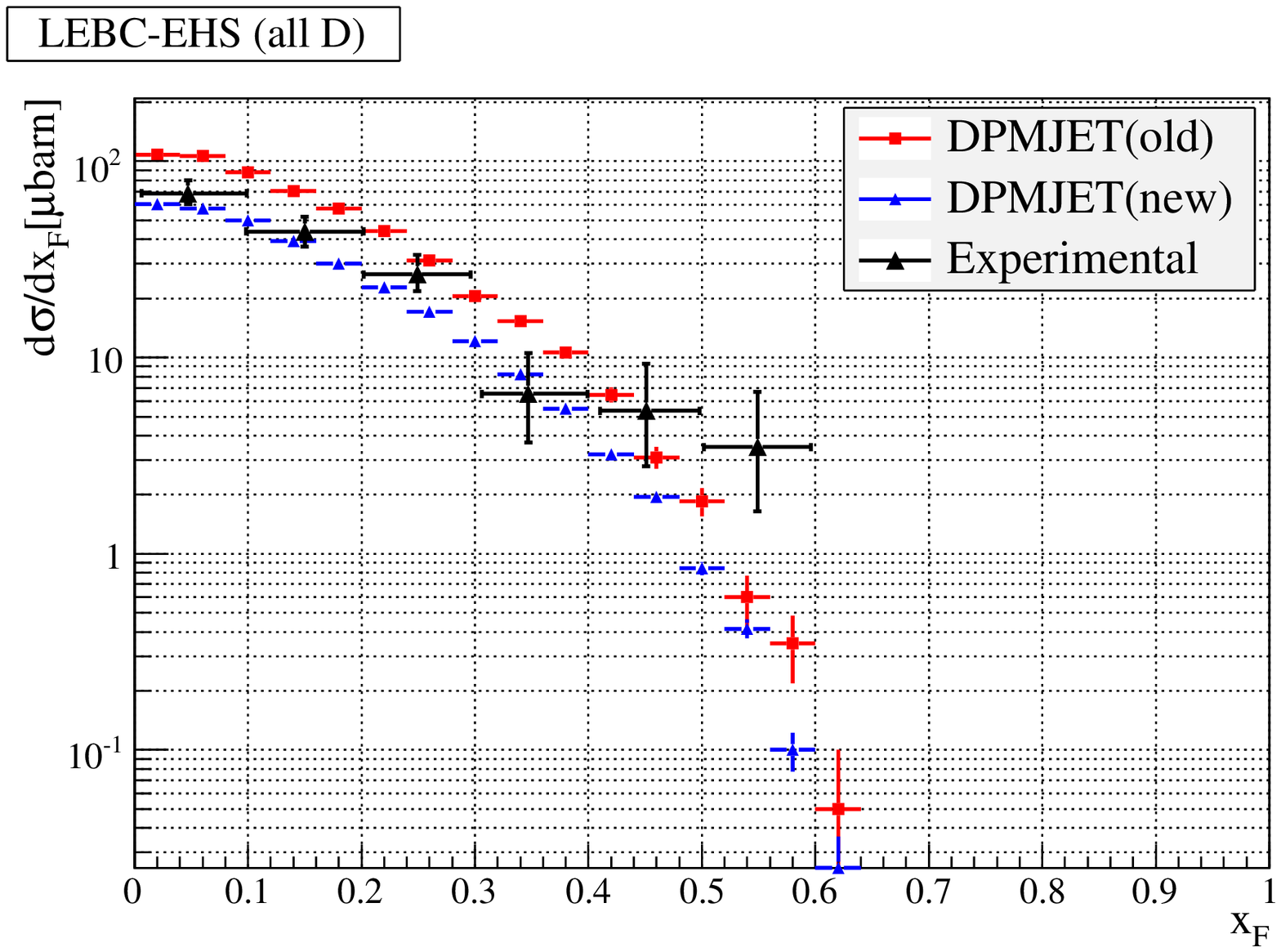}
\includegraphics[width=3in]{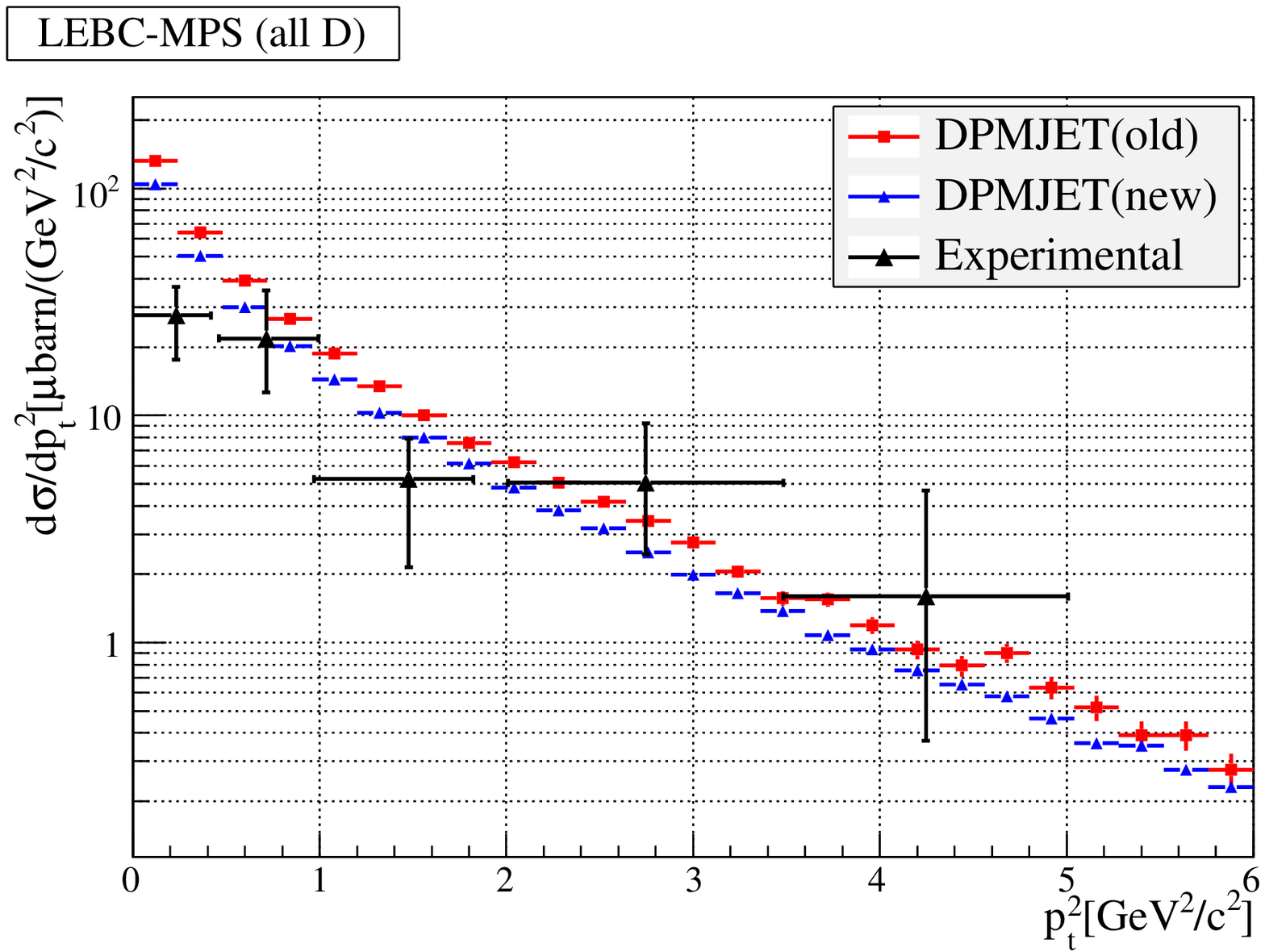}
\includegraphics[width=3in]{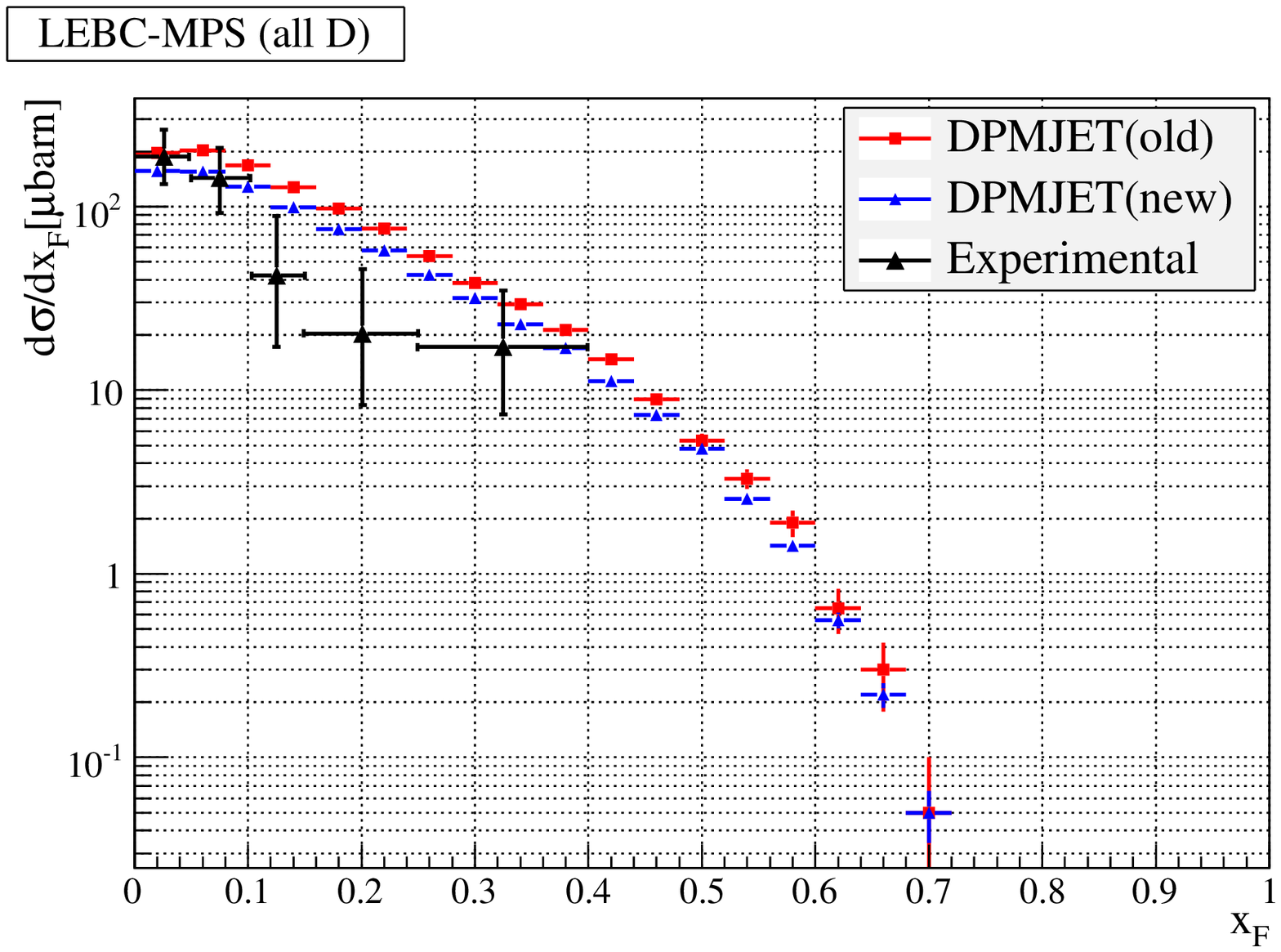}
\caption{LEBC hydrogen bubble chamber measurements, sum over all D-mesons. Values are taken from \cite{AguilarBenitez:1988sb} and \cite{Ammar:1988ta}. Top left: 400GeV $p_{\perp}^{2}$, Top right: 400GeV $x_{F}$, Bottom left: 800GeV $p_{\perp}^{2}$, Bottom right: 800GeV $x_{F}$. In this case, results from the corrected version (cf. Figs. \ref{selex_fig} and \ref{hera_fig}) are identical to DPMJET(new).}
\label{lebc_fig}
\end{figure}

\begin{figure}
\includegraphics[width=3in]{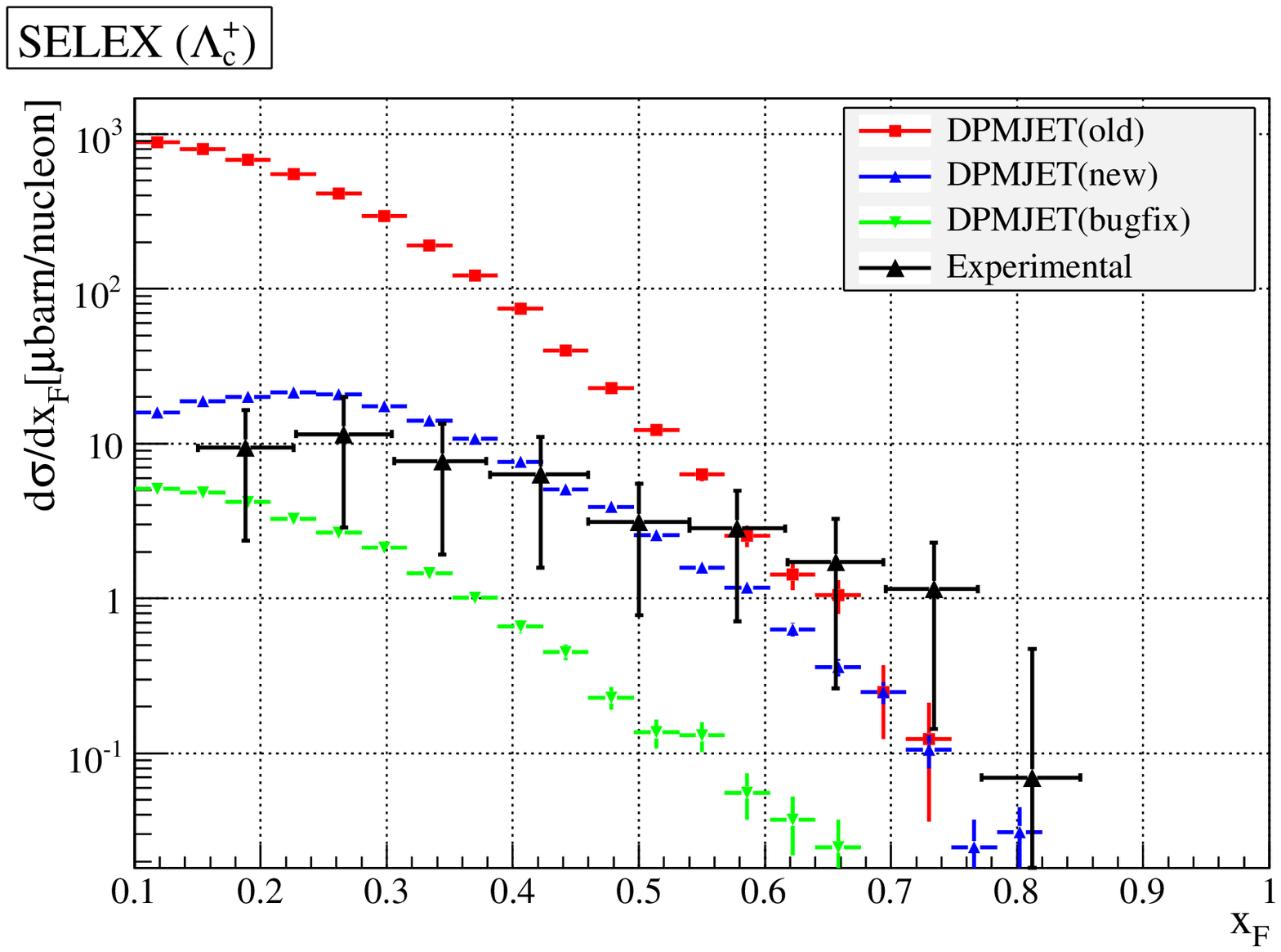}
\includegraphics[width=3in]{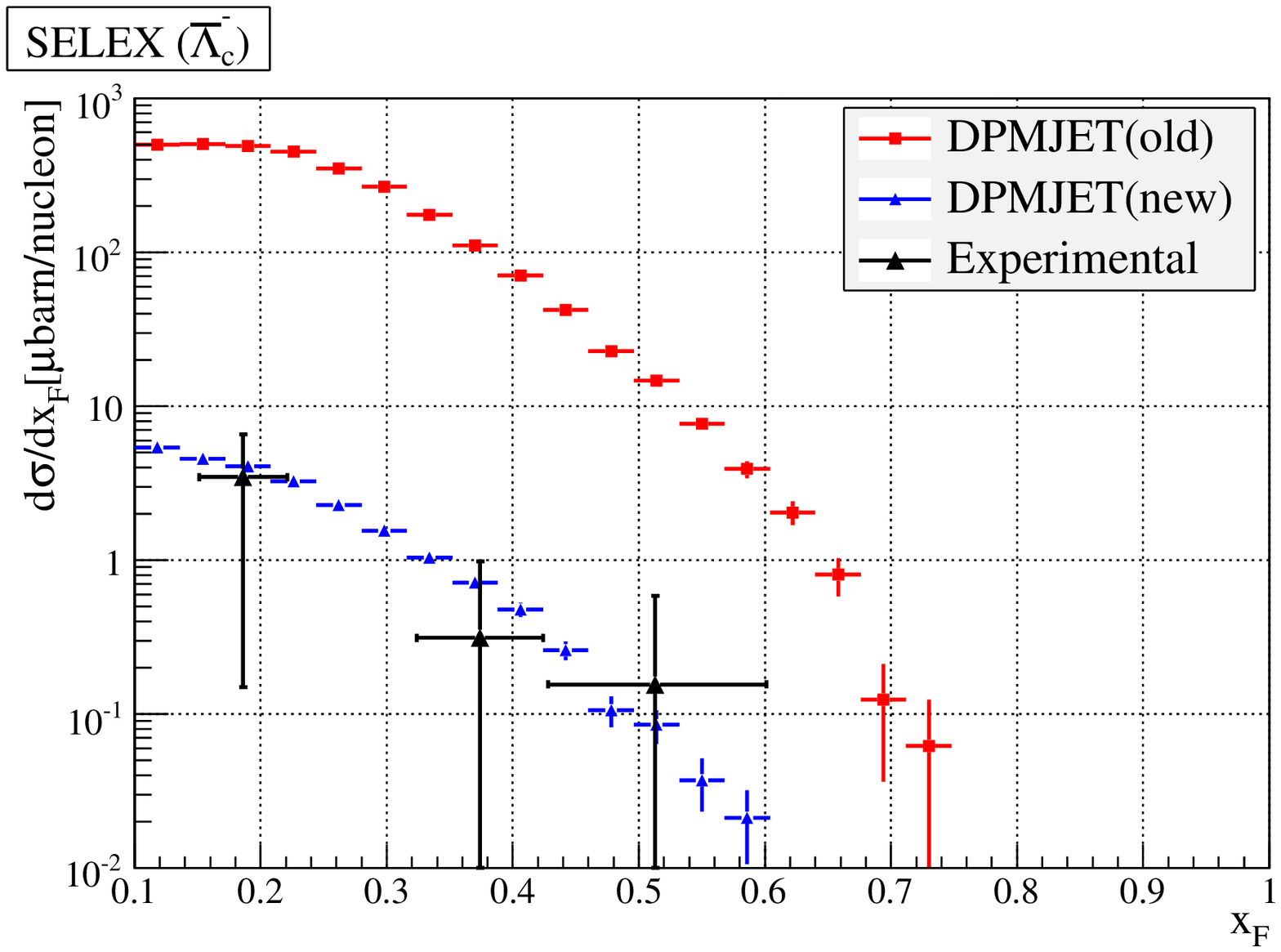}
\includegraphics[width=3in]{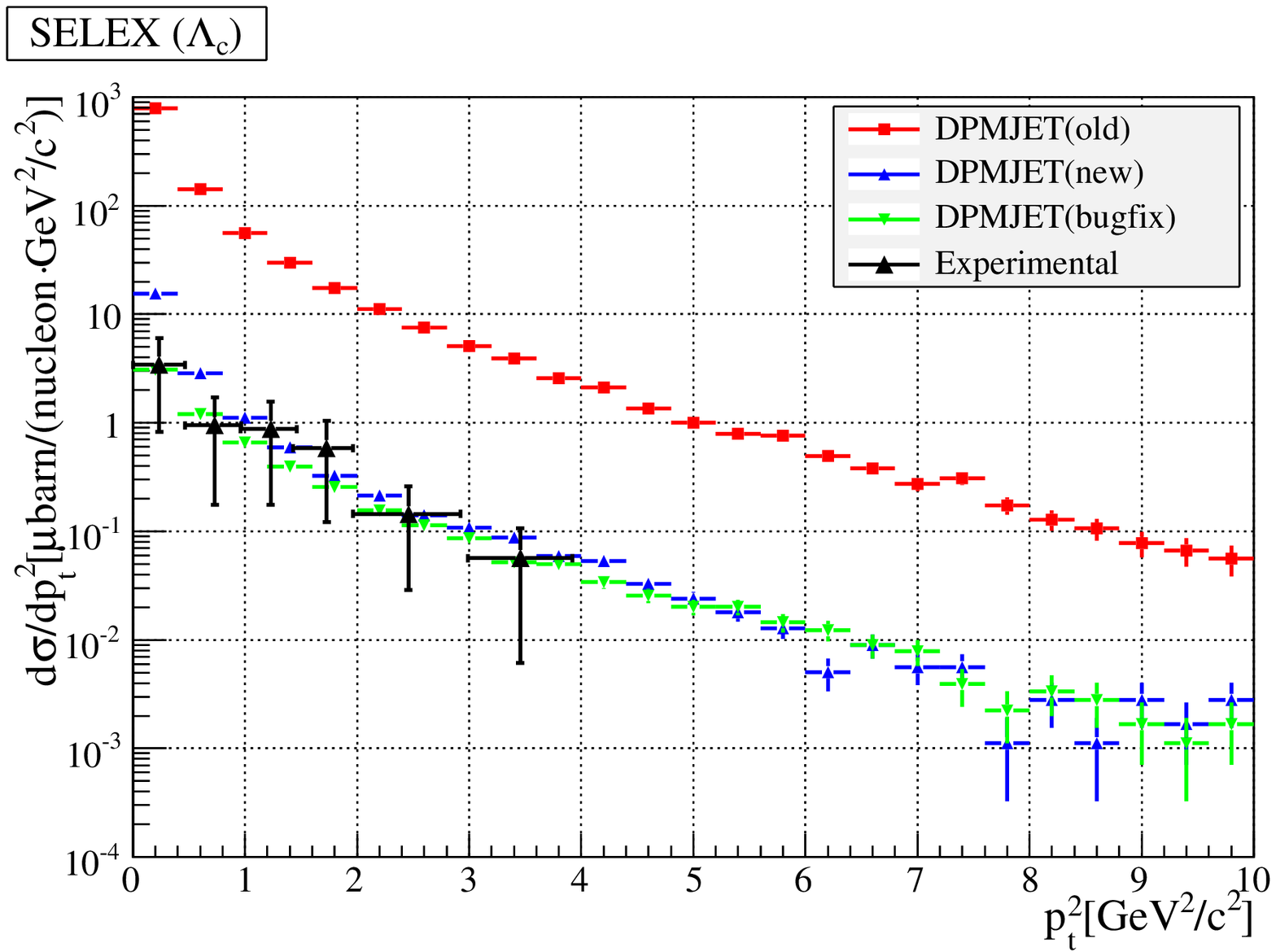}
\includegraphics[width=3in]{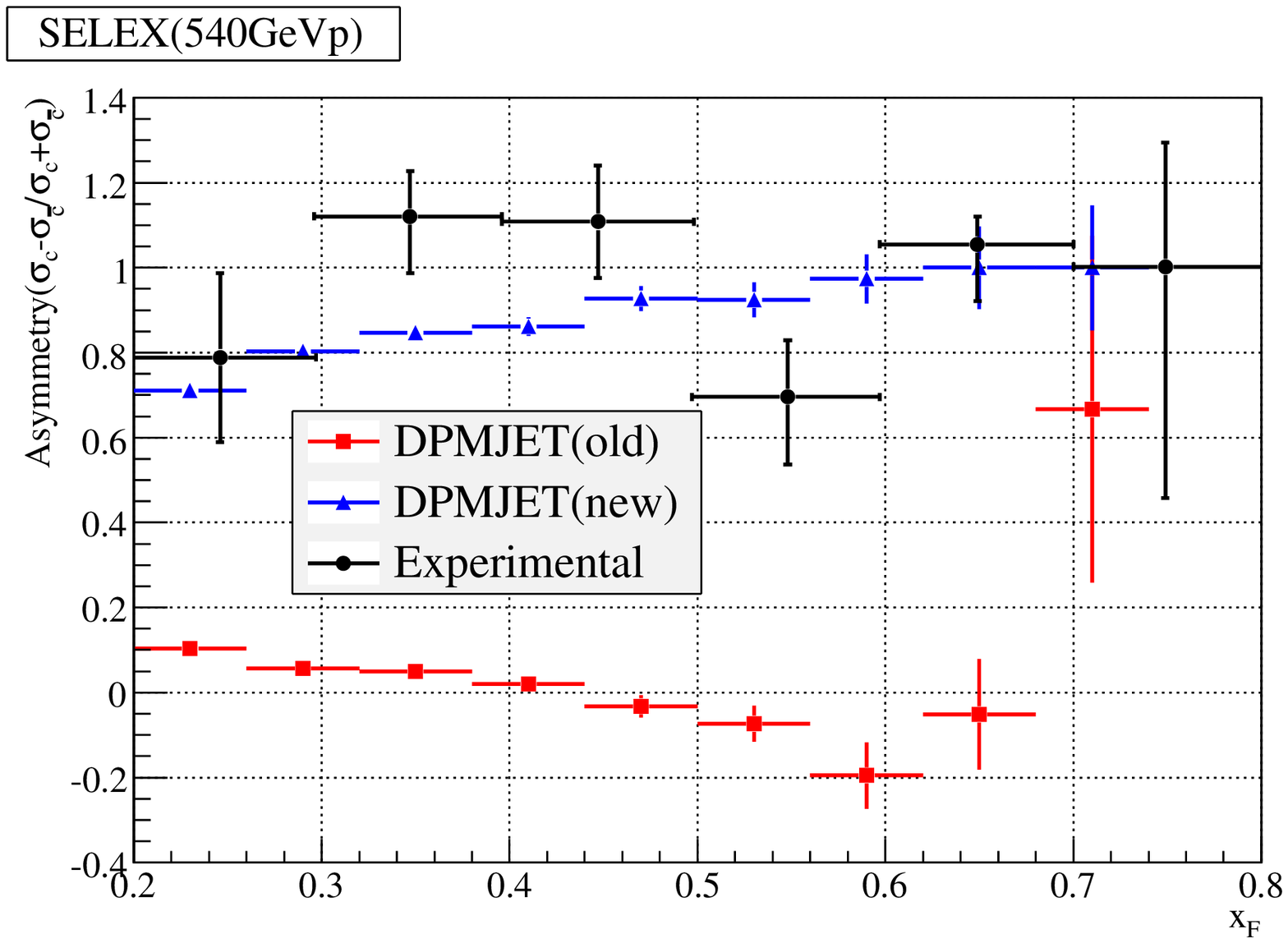}
\caption{SELEX measurements of $\Lambda_{c}$ production parameters and asymmetry for a 540GeV proton beam on a mixed C/Cu target. Top left: $\Lambda_{c}^{+}~x_{F}$, Top right: $\bar{\Lambda}_{c}^{-}~x_{F}$, Bottom left: $p_{\perp}^{2}$ for all $\Lambda_{c}$, Bottom right: Asymmetry between $\Lambda_{c}^{+}$ and $\bar{\Lambda}_{c}^{-}$ production as a function of $x_F$. Asymmetry values higher than one are the result of the likelihood method used to analyze the data. All measurements taken from \cite{Garcia:2001xj}. The error bars have been extended with respect to the original paper to reflect uncertainties in absolute $\Lambda_{c}$ production cross section, for which a value of $7\pm5.25\mu$barn/nucleon was assumed (see text). Shown are the result from the old release version of CORSIKA, the version with the corrected parameter (bugfix) and the newest version in which leading quark effects are implemented. For more information about the different versions, see Section \ref{bugsec}. For the $\bar{\Lambda}_{c}^{-}$, the parameter-corrected result is identical to the new version, in the asymmetry plot it is identical to the old version.}
\label{selex_fig}
\end{figure}

\begin{figure}
\includegraphics[width=3in]{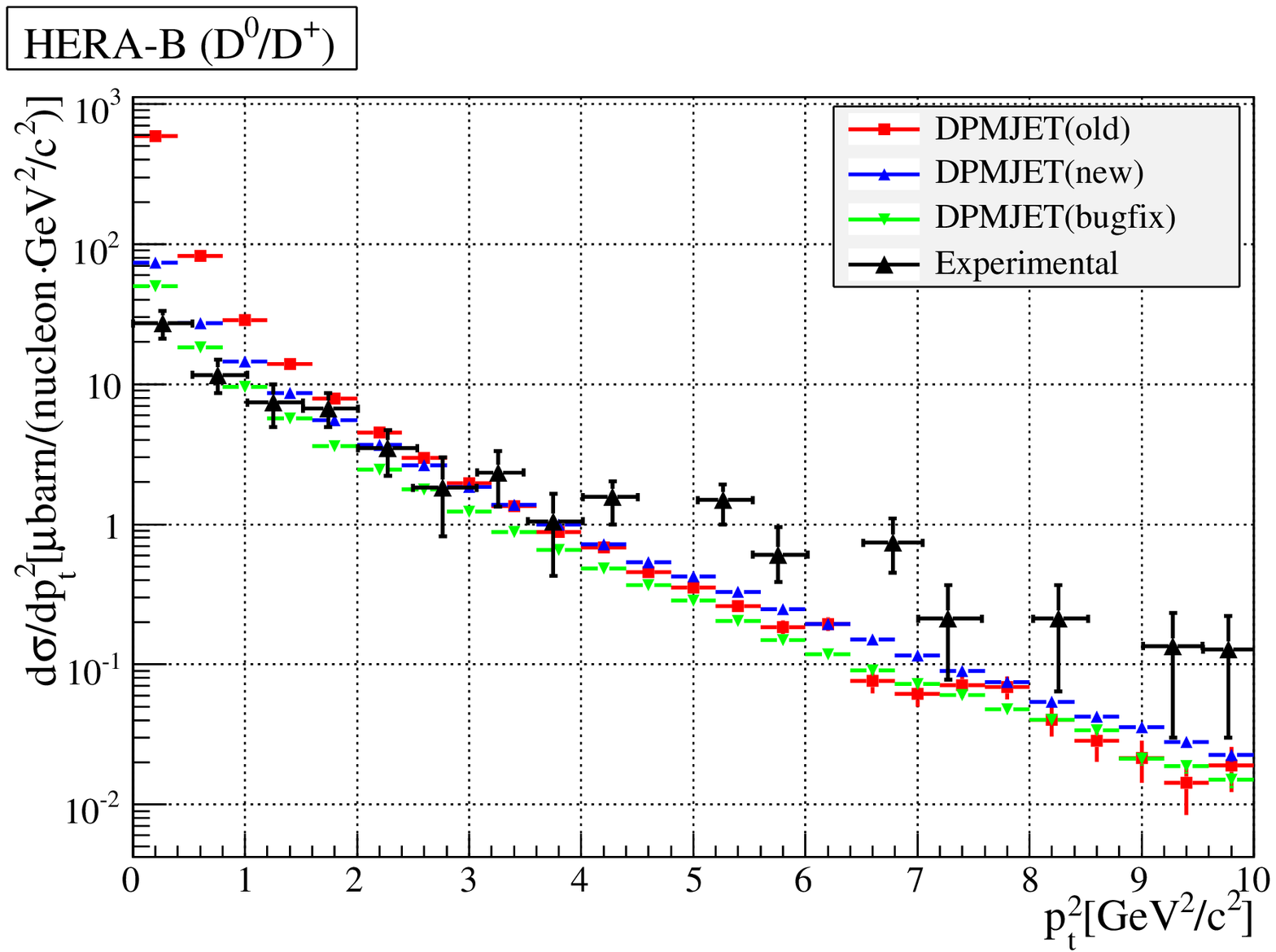}
\includegraphics[width=3in]{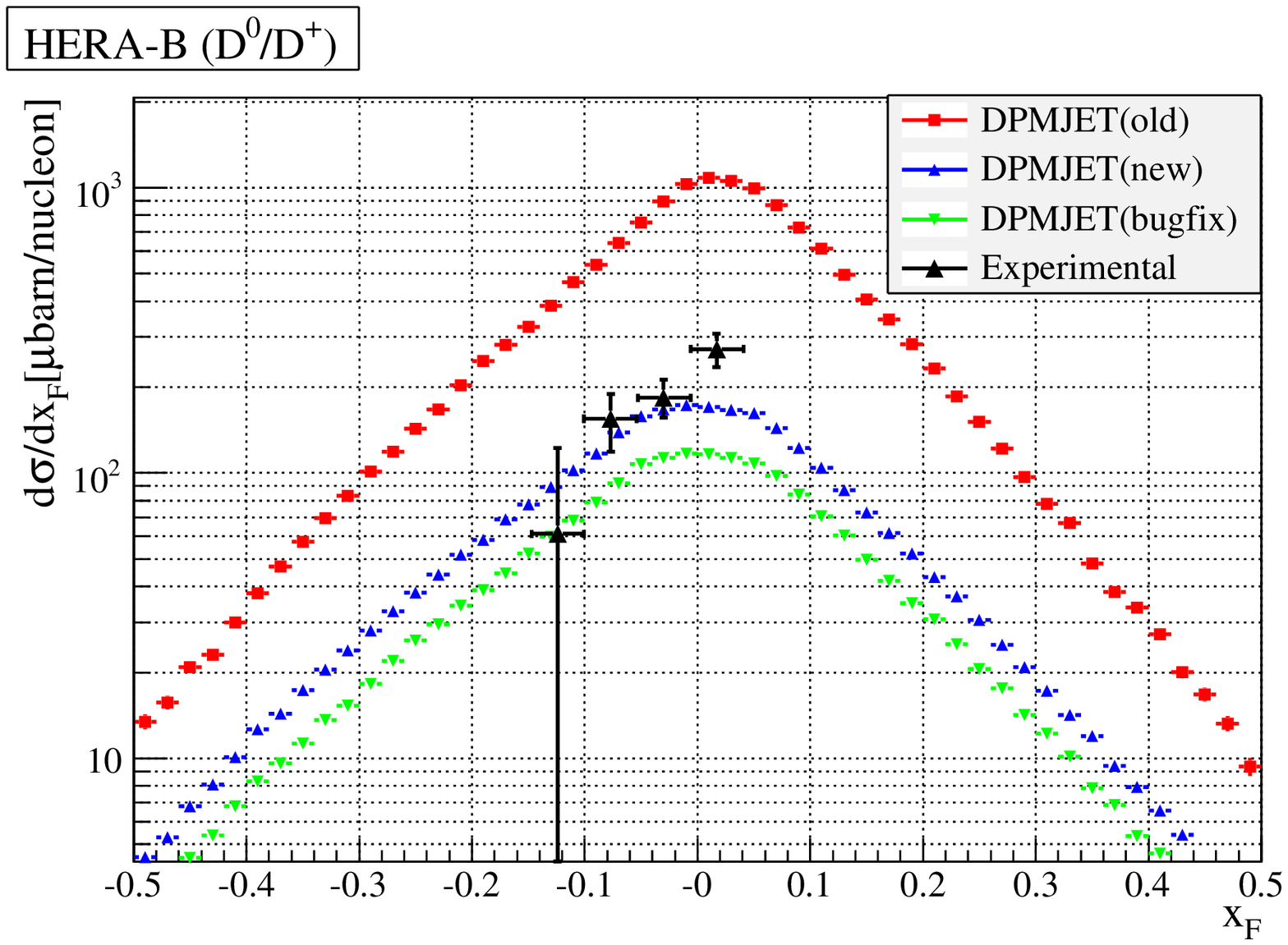}
\caption{HERA-B measurement for D-meson fixed target production using a 920GeV proton beam on C, Ti and W targets. Left: $x_{F}$, Right: $p_{\perp}$. Experimental data are taken from \cite{Abt07}. For explanation of the different DPMJET results, see Figure \ref{selex_fig} and Section \ref{bugsec}.}
\label{hera_fig}
\end{figure}

Precision measurement of differential charm production cross sections at low transverse momentum $p_{\perp}$ and Feynman-x $x_{F}=p_{L}/p_{L_{max}}$ requires comprehensive instrumental coverage of the interaction region, especially in direction of the beam. Since this is difficult to achieve in collider experiments, relevant data are almost exclusively provided by fixed-target setups. In the past, various experiements have been conducted using different types of particle beams. In the following, we will restrict ourselves to the discussion of results that were obtained with proton beams. Even though statistics for proton-nucleus interactions are relatively low compared to those for meson beams \cite{Lourenco:2006vw}, these are the processes that are the most relevant in cosmic ray-induced air showers. Moreover, we consider data in the low $p_{\perp}$ region most relevant for air showers.

Figure \ref{lebc_fig} shows the simplest case, a proton beam impinging on hydrogen nuclei in the LEBC bubble chamber. These results were already used in the first comparison between {\sc dpmjet} and experimental data in 1995 \cite{Battistoni95b}. Taking into account the limited statistics going into the experimental result, agreement between experiment and simulation is satisfactory.

An important aspect of charm hadroproduction is the leading quark effect described in section \ref{sec-selex}, by which production of hadrons which share a common diquark with the incoming projectile is favored. The extent of this effect has been demonstrated by the SELEX colaboration using beams of protons, $\pi^{-}$ and $\Sigma^{-}$. In the forward $x_{F}$ region, production of $\Lambda_{c}^{+}$ was found to be significantly enhanced over its antiparticle for p and $\Sigma^{-}$ beams, whereas for the $\pi^{-}$ beam no asymmetry was observed. Including this effect in the latest vesion of {\sc dpmjet} leads to good agreement with the experimental result, as shown in Figure \ref{selex_fig}.

Since the original publication only gave absolute event numbers \cite{Garcia:2001xj}, and no value for the charm production cross section was published by the collaboration, in Figure \ref{selex_fig} an additional systematic error was added to the experimental result in order to account for uncertainties in the estimate. In \cite{AguilarBenitez:1988sb}, the value for $\Lambda_{c}$ production was found to be
\begin{equation}
1.4\mu barn \leq \sigma(pp\rightarrow \Lambda_{c}^{+} / \bar{\Lambda}_{c}^{-})B(\Lambda_{c} \rightarrow 3 charged) \leq 6.1\mu barn
\end{equation}
which together with the fraction of $\Lambda_{c}$ to total open charm 
\begin{equation}
\sigma(\frac {\Lambda_{c}^{+} / \bar{\Lambda}_{c}^{-}}{c\bar c})=0.28 \pm 0.21
\end{equation}
and the total open charm production cross section from \cite{Lourenco:2006vw}
\begin{equation}
\sigma(pp \rightarrow c \bar c) = 25 \mu barn
\end{equation}
results in an estimate for the charm production at SELEX energies ($540GeV$) as $7\pm5.25 \mu barn/nucleon$. This is consistent with the value obtained from simulating charm production in {\sc dpmjet} ($9.5 \mu barn/nucleon$).

Finally, the simulation was compared to data from HERA-B \cite{Abt07}. Even though the kinematic region covered by the detector was limited to $-0.15<x_{f}<0.05$ and $p_{\perp}<3.5GeV/c$, the result represents the most accurate measurement of open charm production using a proton beam. Figure \ref{hera_fig} shows the comparison with our simulation. Even though there appear to be slight discrepancies in the individual data points, the result from {\sc dpmjet} agrees well with the fit to the data presented by the HERA collaboration in their publication in summer 2007.

SELEX used a mixture of copper and carbon targets, and HERA-B alternated between carbon, titanium and tungsten. In each case, the target thickness was a small fraction of the proton interaction length. To simplify the simulation, in {\sc dpmjet} nitrogen nuclei were used as target material. This is legitimized by the fact that, as noted above, collision scaling was verified both in experiment and simulation. The individual diagrams have been scaled to represent the cross section values per nucleon.

\section{Conclusion}

Charm production in {\sc dpmjet} has been checked against experimental data from both fixed target and collider experiments. Since the primary purpose of {\sc dpmjet} is simulation of cosmic ray-induced air showers, the comparison was restricted to proton-nucleon interactions at low $p_{\perp}$, which however represent the by far dominant contribution to charm production. The total open charm cross section is consistent with experimental values, even though systematic errors on measurements are very large, especially at high energies. 

Comparison of simulation to differential distributions from fixed-target experiments show reasonable agreement. Since 1995, new data have become available that show hard collisions playing a much larger role than previously assumed. Those results, most notably from the SELEX collaboration concerning charmed baryon asymmetries, but also collision scaling in $c \bar c$ production, were incorporated into the code.

In summary, charm production in {\sc dpmjet} has been updated reflecting the latest experimental results and is now consistent with all available measurements. All simulations have been performed with the {\sc dpmjet} version integrated in the CORSIKA air-shower simulation package. It thus allows for the first time to generate Monte-Carlo simulations of the prompt component in cosmic ray air showers. The precise effect of prompt contributions on muon and neutrino fluxes, as well as a study of meson-nucleon collisions, will be subject of a later paper.




\end{document}